\documentclass[aip,apl,reprint,amsmath]{revtex4-1}

\usepackage{graphicx}
\usepackage{dcolumn}
\usepackage{amsmath}
\usepackage{bm}
\usepackage{float}

\newcommand {\urss}[1]{\ensuremath{_{\mathrm{#1}}}}
\newcommand{\nist}{\affiliation{National Institute of Standards and
    Technology, 325 Broadway MS\,817.03, Boulder, Colorado 80305, USA}}
\newcommand{\cu}{\affiliation{University of Colorado,
    Boulder, Colorado 80309 USA}}
\newcommand{\nistauthor}[1]{\author{#1}\nist}
\newcommand{\cuauthor}[1]{\author{#1}\nist\cu} 
\newcommand{\phiCDM}{$\Phi$-CDM}

\newcommand{\mat}[1]{\ensuremath{\mathbf #1}}   

\begin{document}

\title{
  Code-division multiplexing for x-ray microcalorimeters
}

\nistauthor{G.M. Stiehl}

\cuauthor{W.B. Doriese}

\author{J.W. Fowler} \email[Electronic mail: ]{joe.fowler@nist.gov}
\nist\cu

\nistauthor{G.C. Hilton}

\nistauthor{K.D. Irwin}

\nistauthor{C.D. Reintsema}

\nistauthor{D.R. Schmidt}

\nistauthor{D.S. Swetz}

\nistauthor{J.N. Ullom}

\nistauthor{L.R. Vale}

\date{\today}

\begin{abstract}


  We demonstrate the code-division multiplexed (CDM) readout of eight transition-edge
  sensor microcalorimeters.  The energy resolution is 3.0\,eV (full width at
  half-maximum) or better at 5.9\,keV, with a best resolution of 2.3\,eV and a mean
  of 2.6 eV over the seven modulated detectors. The flux-summing CDM system is
  described and compared with similar time-division multiplexed (TDM) readout.  We
  show that the $\sqrt{N\urss{pixels}}$ multiplexing disadvantage associated with TDM
  is not present in CDM\@.  This demonstration establishes CDM as both a simple route
  to higher performance in existing TDM microcalorimetric experiments and a long-term
  approach to reaching higher multiplexing factors.

\end{abstract}


\maketitle

The transition-edge sensor\cite{Irwin2005} (TES) is an established calorimetric
detector whose applications include x-ray astronomy,\cite{kilbourne2008} gamma-ray
spectroscopy for analysis of nuclear materials,\cite{Dor07} and probing molecular
dynamics through x-ray absorption spectroscopy.\cite{Bressler04,Uhlig12} These
applications demand ever-larger arrays of detectors to increase photon throughput.
Because TES microcalorimeters are operated at sub-Kelvin temperatures, the reduction
of power dissipation and wire count through multiplexing is crucial. Ammeters made
from superconducting quantum interference devices (SQUIDs) are widely used to read out
TESs due to their low noise, low impedance, low power dissipation, and high
bandwidth.

A typical multiplexed array consists of multiple independent amplifier channels each
reading out $N$ detectors.  The two most mature multiplexing techniques now in use
for TESs are time-division multiplexing (TDM)\cite{kilbourne2008} and MHz-band
frequency-division multiplexing (FDM).\cite{yoon:371} Neither is ideal for
microcalorimetry. In an $N$-row TDM multiplexer, the SQUID noise aliased into the
signal band grows as $\sqrt{N}$, a consequence of inefficient use of the readout
bandwidth.\cite{Dor06} This noise limits multiplexers for high-resolution x-ray or
gamma-ray microcalorimeters to tens of detectors per amplifier channel.  FDM avoids
the $\sqrt{N}$ noise penalty but has its own limitations when operated in the MHz
range, including physically large filter components,\cite{yoon:371} and the
degradation of sensor resolution by ac biasing.\cite{gottardi2011}

A third multiplexing technique, code-division multiplexing (CDM),\cite{Irwin2010}
is being developed at NIST in two distinct configurations: CDM through current
summation\cite{niemack:163509} ($I$-CDM), and CDM through flux
summation\cite{Fowler11} (\phiCDM).  In this letter, we report on an eight-element
array of TES microcalorimeters read out through \phiCDM.  The energy resolution
at 5.9 keV averaged 2.6\,eV full-width at half-maximum (FWHM) in the seven
modulated detectors.  We summarize the \phiCDM\ design and compare its noise with
that of a similar TDM system.

\begin{figure}[ht!]
\includegraphics[width=\columnwidth]{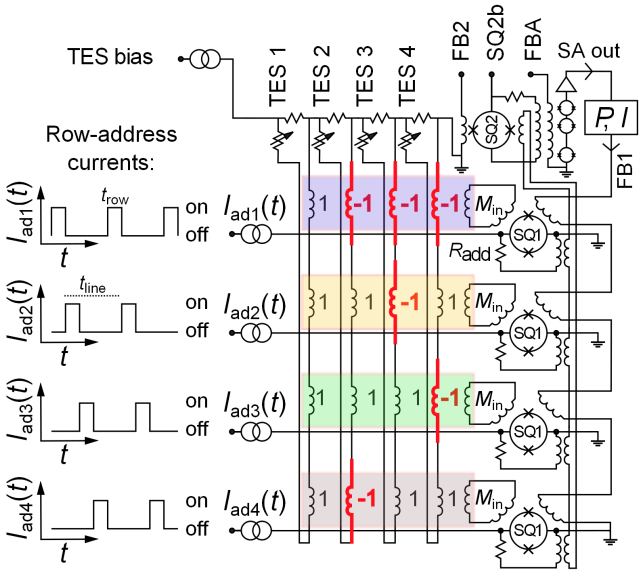}
\caption{\label{fig:FluxSumSchematic}
  (Color online) A four-row implementation of code-division multiplexing by flux
  summation (\phiCDM).  The TESs are dc-biased and thus on at all times.  The current
  signal from TES $j$ inductively couples to all four first-stage SQUID amplifiers
  (SQ1) with coupling polarity defined by column $j$ in the modulation matrix $W_{4}$
  (Equation~\ref{eq:Walsh 4x4_matrix}).  Oppositely oriented inductors (red/bold)
  produce a negative coupling polarity. Each row of inductors (shaded boxes) is
  transformer-coupled to one SQ1. Rows of SQ1s are operated with a standard TDM
  protocol (see Ref.~\citenum{Rei03}): the rows are activated sequentially via
  $I_{\mathrm{ad}k}$, so the signal from one SQ1 at a time passes to a second-stage
  SQUID (SQ2). The output of SQ2 is routed to a 100-SQUID, series-array amplifier and
  then to room-temperature electronics. To keep the three-stage SQUID amplifier in
  its linear range, the multiplexer is run as a flux-locked loop
  (Ref.~\citenum{Rei03}). The series array output (SA-out) is digitally sampled; a
  flux-feedback signal FB1 is then applied inductively to each SQ1 to maintain SA-out
  at a constant value.}
\end{figure}

\begin{figure}
\includegraphics[width=3.25in]{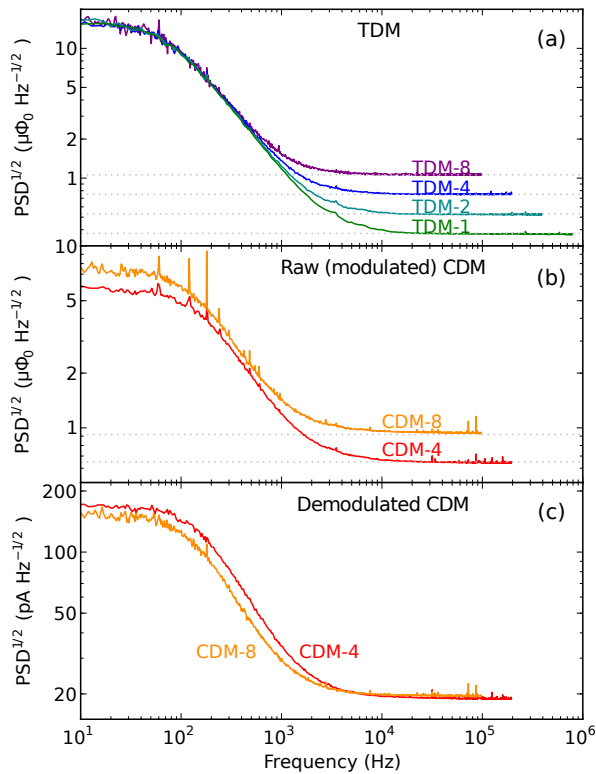}
\caption{\label{Noise}
 (Color online) The scaling of SQUID-amplifier noise in TDM and \phiCDM.  Noise was
  measured at 85~mK, with TESs superconducting to emphasize the amplifier noise
  (rather than TES noise) at high frequencies.  The Johnson-noise contribution from
  the TES shunt resistor dominates below 1\,kHz.  The $\tau=L/R$ time constant of the
  shunt resistance and inductance in the TES bias loop causes the Johnson noise to
  roll off above 100\,Hz.  At high frequencies, the SQUID-amplifier noise is
  dominant.  All measurements used $t\urss{row}=640$~ns and a 2.5\,MHz, one-pole $RC$
  filter before the digitizer.
  {(a)} Noise from a single SQUID channel, referred to the first-stage SQUID, when
  read out with one, two, four, or eight TDM rows.  Dotted lines show the single-row,
  high-$f$ noise level (0.37~$\mu\Phi\urss{0}/\surd$Hz) multiplied by successive powers of
  $\sqrt{2}$.  Due to aliasing, TDM amplifier noise grows with the number of rows as
  $\sqrt{N}$ (see Ref.~\citenum{Dor06}).
  {(b)} Noise in four- and eight-channel CDM readout.  The signals, which have not
  been demultiplexed via the Walsh matrix, are referred to the first-stage SQUID\@.
  Lines are seen at the 60\,Hz power line frequency and its harmonics.
  The dotted lines show the CDM-4, high-$f$ noise level (0.65~$\mu\Phi\urss{0}/\surd$Hz)
  multiplied by   1 and $\sqrt{2}$.  As in TDM, the aliased SQUID noise scales as
  $\sqrt{N}$.
  {(c)} Demodulated noise, referred to the TES current, in four- and eight-channel
  CDM.  Both approach 19\,pA/$\surd$Hz at high frequencies.  We omit the unswitched
  channel from the average, making the 60\,Hz line no longer visible.
}
\end{figure}

We have fabricated \phiCDM\ multiplexers to read out arrays of 4, 8 and 16 detectors.
Figure~\ref{fig:FluxSumSchematic} depicts a four-detector array and explains its
operation.  The TES signals are encoded through a Walsh basis set\cite{Walsh1923}
defined by the polarity of lithographically patterned inductive traces. The
four orthogonal combinations of the signals are read out in sequence. The encoding
matrices used in the 4- and 8-detector designs are
\begin{equation} \label{eq:Walsh 4x4_matrix}
W_{4} \equiv \left( \begin{array}{rrrr}
 1 & -1 & -1 & -1 \\
 1 & 1 & -1 & 1 \\
 1 & 1 & 1 &  -1 \\
 1 & -1 & 1 & 1
\end{array} \right), 
\end{equation}
\begin{equation} \label{eq:Walsh 8x8_matrix}
\mathrm{and}\ W_{8} \equiv \left( \begin{array}{rrrrrrrr}
 1 & -1 &  1 & -1 & -1 &  1 & -1 & -1 \\
 1 &  1 &  1 &  1 & -1 & -1 & -1 &  1 \\
 1 &  1 &  1 & -1 &  1 & -1 &  1 & -1 \\
 1 & -1 &  1 &  1 &  1 &  1 &  1 &  1 \\
 1 & -1 & -1 &  1 &  1 & -1 & -1 & -1 \\
 1 &  1 & -1 & -1 &  1 &  1 & -1 &  1 \\
 1 &  1 & -1 &  1 & -1 &  1 &  1 & -1 \\
 1 & -1 & -1 & -1 & -1 & -1 &  1 &  1
\end{array} \right),
\end{equation}
where matrix columns represent TES detectors, and the rows represent readout rows.
The encoding matrix $W_4$ or $W_8$ gives the coupling polarity between the signal
from each detector and each readout row.  (Viewed in this way, a TDM system uses
the identity matrix for encoding: $W\urss{TDM}\equiv\mat{I}$.)  The Walsh code
switches the polarity of each TES but the first, eliminating sensitivity in
demodulated data to any amplifier drift or pickup occurring after the modulation
(e.g., the 60\,Hz power-line harmonics visible in Figure~\ref{Noise}b but absent from
\ref{Noise}c).  In the most demanding applications, the single unswitched input could
be used without a TES as a ``dark SQUID'' noise monitor.  Other than the multiplexer
chips, all the hardware required by \phiCDM\ (SQUID series arrays, wire-bonded cryogenic
circuit boards, and room temperature electronics\cite{Rei03}) is directly
interchangeable with TDM\@. Existing TDM systems thus need no
modifications to their firmware or to the data acquisition software to be ``drop-in
compatible'' with \phiCDM.  The analysis software must be enhanced, however, to
demodulate the $N$ channels of raw data into the detector timestreams.

\begin{figure}[t]
\includegraphics[width=3.25in]{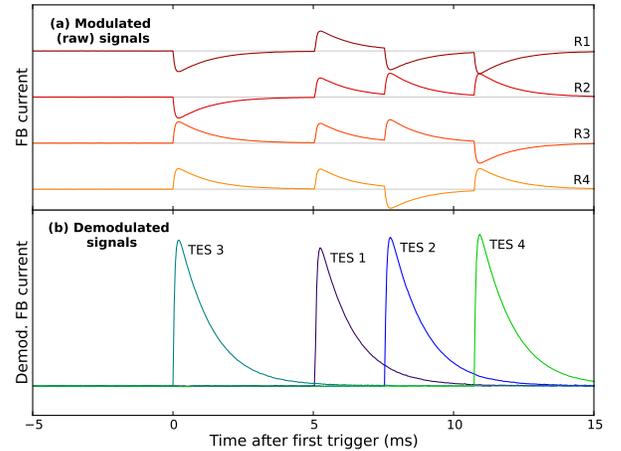}
\caption{\label{fig:CDM4x4pulses}(Color online)
  Example raw and demodulated data from four detectors in a single 20\,ms period.
  {(a)} The raw, encoded outputs, R$k$, from the SQ1 in four-detector \phiCDM\ (with
  vertical offsets for clarity).  The SQ1 outputs correspond to rows 1--4
  in Equation~\ref{eq:Walsh 4x4_matrix}.  Manganese fluorescence x-rays struck TESs
  3, 1, 2, and 4 at 0, 5, 7, and 11\,ms.
  {(b)} The same data demodulated by application of
  $W^{-1}_{4}$ to show the per-detector signal currents.  The signal-to-noise is 
  too high for the noise to be seen in this example.}
\end{figure}

An example of Walsh-encoded and -decoded data is shown in
Figure~\ref{fig:CDM4x4pulses}.  Four photons arrive during a 20\,ms window on a
four-detector \phiCDM\ array.  The top panel shows the encoded signal recorded by
each first-stage SQUID as the four detectors each absorb x-rays.  The photons strike
TES 3, 1, 2, then 4; the encoded pulse polarities therefore reflect columns of
Equation~\ref{eq:Walsh 4x4_matrix} in the same order.  The bottom panel shows the
reconstructed signal currents in the individual TESs over the same 20\,ms.

Because the encoding matrices are defined by lithography on the multiplexer chip,
details of the inductor layout and other on-chip sources of cross-talk produce
unequal couplings between the detectors and SQUIDs. The encoding matrices given in
Equations~\ref{eq:Walsh 4x4_matrix} and \ref{eq:Walsh 8x8_matrix} are therefore only
idealizations; we have measured the true encodings to depart from the ideal at the
1\,\% to 2\,\% level in 4, 8, and 16-detector CDM multiplexers.  Measurements on a partial
32-detector multiplexer suggest its non-uniformity will be at the same level.  A
correction computed in offline analysis\cite{Fowler11} can reduce this imbalance to
levels below 0.1\,\%. Corrected decoding matrices are used for all \phiCDM\ data in
this letter.  We find this correction to be stable over at least several weeks,
allowing it to be measured once and then applied in real-time data analysis.  The
demodulation requires computation scaling as $N^2$  per data sample for $N$
multiplexed detectors.  Our experience shows that the computational burden will
not prevent scaling the technique up to at least $N\approx 100$.

A further linear arrival-time correction is applied,\cite{Fowler11} though it makes
no significant difference in the present observations.  At higher $N$, where the time
between successive samples grows longer, the correction would help by reducing the
dependence of demodulated pulse shapes on the arrival time.  The correction will also
be important to reduce cross-talk effects in future data sets with larger $N$ and
with higher photon rates.

\begin{figure}
\includegraphics[width=3.25in]{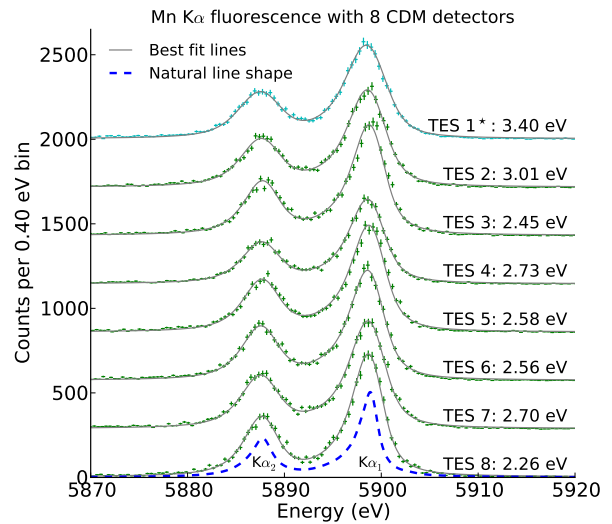}
\caption{\label{fig:TempSpectra}
  (Color online) Mn K$\alpha$ x-ray fluorescence spectra measured separately by eight
  TES x-ray calorimeters read out with \phiCDM.  Spectra are offset vertically for clarity.
  These data have been analyzed with corrected Walsh codes and a linear arrival-time
  correction, and a Gaussian energy resolution has been fit, techniques described
  previously in Ref.~\citenum{Fowler11}.  All detectors have multiplexed energy
  resolution better than 3\,eV except for TES 1$^\ast$---the only detector subject to
  low-frequency noise pickup in the SQUID amplifier chain.  The \phiCDM\ resolution
  matches or exceeds that found with equivalent TESs read out by TDM\@.
}
\end{figure}

In a TDM channel of $N$ detectors, the wide-band SQUID amplifier noise level (at fixed
sampling rate) scales as $\sqrt{N}$ due to aliasing (Figure~\ref{Noise}a).  The
relation between the detector current noise $I\urss{Namp(TES)}$ and SQUID flux
noise $\Phi\urss{Namp(SQ1)}$ in TDM is\cite{Dor06}
\begin{equation}\label{eq:Current noise to Flux noise}
I\urss{Namp(TES)}=\Phi\urss{Namp(SQ1)}\sqrt{\pi N}/M\urss{in}.
\end{equation}
For large $N$, the coupling mutual inductance $M\urss{in}$ of the TES signal to the
SQ1 amplifier must be increased to compensate for the higher amplifier noise.
$M\urss{in}$ is limited on the high end by the dynamic range of the SQ1 when tracking
the steep leading edge of photon pulses.



CDM has the advantage of sampling all detectors at all times, while TDM samples
each only $1/N$ of the time.  This means that amplifier bandwidth is used much more
efficiently in CDM\@.  In practice, this works as follows.  The encoded SQ1 signals
(Figure~\ref{Noise}b) suffer the same $\sqrt{\pi N}$ multiplex disadvantage as in TDM.
In decoding, however, TES signals average coherently, while the $N$ samples of
amplifier noise average incoherently.  The demodulated TES signal and amplifier noise
are therefore independent of multiplexing factor (Figure~\ref{Noise}c).  \phiCDM\ thus
allows $M\urss{in}$ to remain low at large $N$ without increasing the amplifier noise
as referred to TES current.  Low mutual inductance, in turn, increases the effective
dynamic range of the SQUID amplifiers and makes the system more robust with fast TESs
and in the face of high pulse rates.

Figure~\ref{fig:TempSpectra} shows Mn K$\alpha$ fluorescence spectra measured by
eight TES x-ray detectors read out with \phiCDM. All detectors (besides the
unswitched TES~1) achieved 3.0\,eV FWHM energy resolution or better at 5.9 keV.  The
mean resolution of 2.6\,eV is better than the best previous multiplexed TES
measurement at this energy.\cite{kilbourne2008} Count rates in these data are low
(approximately 5\,Hz per detector), but we anticipate operating with much higher
rates in the near future.


We view the demonstration of \phiCDM\ presented here as important for two
reasons.  First, \phiCDM\ chips are drop-in compatible with existing 32-row TDM
systems but have higher performance.  They offer an immediate path to the
kilopixel-scale arrays of high-resolution TES microcalorimeters desirable in
applications like synchrotron science and the proposed Athena satellite.
Second, the \phiCDM\ system provides a bridge to the eventual development of $I$-CDM,
in which the rapid alternation of SQUID switches replaces transformer
windings as the mechanism for encoding TES signals.  An $I$-CDM multiplexer could scale
to hundreds of detectors per amplifier channel,\cite{Irwin2010} eventually enabling even
megapixel-scale arrays.

NASA grant NNG09WF27I and an American Recovery and Reinvestment Act Fellowship
to JF supported this work.  Contribution of NIST, not subject to copyright.

\bibliography{CDM_08_26}

\end{document}